\documentclass[twocolumn,showpacs,aps,floatfix]{revtex4}
\usepackage{times}
\usepackage{epic}
\usepackage{eepic}
\usepackage{amssymb}
\usepackage{amsmath}
\usepackage{amsbsy}
\usepackage{amsfonts}
\usepackage{amsthm}
\newcommand{\rr}{\mathbb R}
\newcommand{\cc}{\mathbb C}
\newtheorem{theo}{Theorem}

\newcommand{\ket}[1]{\left\vert #1 \right\rangle}

\newcommand{\project}[2]{\left\vert #1 \right\rangle\left\langle #2 \right\vert}
\newcommand{\mean}[1]{\left\langle #1 \right\rangle}

\newcommand{\one}{\mbox{\large{$1 \hspace{-0.95mm}  {\bf l}$}}}
\newcommand{\EP}{\mathtt{EP}}
\newcommand{\SWAP}{\mathtt{SWAP}}

\begin{document}

\title{Optimal Quantum Circuits for General Two-Qubit Gates}

\author{Farrokh Vatan} \email{Farrokh.Vatan@jpl.nasa.gov}
\author{Colin Williams} \email{Colin.P.Williams@jpl.nasa.gov}
\affiliation{Jet Propulsion Laboratory, California Institute of
Technology, 4800 Oak Grove Drive, Pasadena, CA 91109--8099}

\date{\today}
\pacs{03.67.Lx, 03.65.Fd, 03.65.Ud}

\begin{abstract}
In order to demonstrate non-trivial quantum computations
experimentally, such as the synthesis of arbitrary entangled
states, it will be useful to understand how to decompose a desired
quantum computation into the shortest possible sequence of
one-qubit and two-qubit gates. We contribute to this effort by
providing a method to construct an {\em optimal} quantum circuit
for a general two-qubit gate that requires at most 3 CNOT gates
and 15 elementary one-qubit gates. Moreover, if the desired
two-qubit gate corresponds to a purely real unitary
transformation, we provide a construction that requires at most 2
CNOTs and 12 one-qubit gates. We then prove that these
constructions are optimal with respect to the family of CNOT,
$y$-rotation, $z$-rotation, and phase gates.
\end{abstract}

\maketitle

\section{Introduction}

It is known that any $n$-qubit quantum computation can be achieved
using a sequence of one-qubit and two-qubit quantum logic gates
\cite{nielsen-chuang,barenco}. However, even for two-qubit gates,
finding the {\em optimal} circuit with respect to a particular
family of gates is not easy \cite{divincenzo}. This is unfortunate
because, at the current time, quantum computer experimentalists
can only achieve a handful of gate operations within the coherence
time of their physical systems \cite{braunstein}. Without a
procedure for optimal quantum circuit design, experimentalists
might be unable to demonstrate certain quantum computational
milestones even though they ought to be within reach. For example,
a current experimental goal is the synthesis of any two-qubit
entangled state \cite{arda}. Although it is known, in principle,
how to synthesize any such state \cite{colin}, the resulting
quantum circuits can be suboptimal, requiring excessive numbers of
CNOT gates, if done injudiciously \cite{cybenko}. The current
solution to this problem uses rewrite rules to recognize and
eliminate redundant  gates. However, a better solution would be to
perform optimal design from the outset.

In this paper we give a procedure for constructing an optimal
quantum circuit for achieving a general two-qubit quantum
computation, up to a global phase, which requires at most 3 CNOT
gates and 15 elementary one-qubit gates from the family
$\{R_y,R_z\}$. We prove that this construction is {\em optimal},
in the sense that there is no smaller circuit, using the same
family of gates, that achieves this operation. In addition, we
show that if the unitary matrix corresponding to our desired gate
is purely real, it can be achieved using at most 2 CNOT gates and
12 one-qubit gates.

A flurry of recent results on gate-count minimization for general
two-qubit gates, report similar findings to us. Vidal and Dawson
proved that 3 CNOTs are sufficient to implement a general
$U\in\mbox{\bf SU}(4)$ and that two-qubit controlled--$V$
operations require at most 2 CNOTs \cite{vidal}. Vatan and
Williams proved that any $U\in\mbox{\bf SU}(4)$ requires at most 3
CNOTs, and 16 elementary one-qubit $\{R_y,R_z\}$ gates, that any
$U\in\mbox{\bf SO}(4)$ (i.e., real gate) requires at most 2 CNOTs
and 12 one-qubit $\{R_y,R_z\}$ gates, and that these constructions
are optimal \cite{vatan}. Later, Shende, Markov, and Bullock
reported similar results on circuit complexity for $U\in\mbox{\bf
SU}(4)$, and specialized the complexity bounds depending on which
families of one-qubit gates were being used \cite{shende}.
Fundamentally, all these results rest upon the decomposition of a
general $U\in\mbox{\bf SU}(4)$ given in \cite{khaneja,kraus-cirac}
and used in the GQC quantum circuit compiler \cite{bremner}.

The remainder of the paper is organized as follows. After
introducing some notation in Section~\ref{notation-section}, 
we discuss the {\em magic} basis
\cite{khaneja} in Section~\ref{magic-basis-sec}, 
and prove (in Theorems~\ref{magic-gate-theorem} and 
\ref{magic-gate-2-theorem}) its most important
property, namely, that real entangling two-qubit operations become
non-entangling in the magic basis. We also prove (via the circuit
shown in FIG.~\ref{magic-fig}, first introduced in \cite{vatan}) that the magic
basis transformations require at most {\em one} CNOT to implement
them explicitly. This is in contrast to Fig. 3 in
\cite{bullock-markov}, which required three CNOTs. It turns out
that this compact quantum circuit for the magic basis
transformation is the cornerstone of our subsequent constructions
for generic two-qubit gates, and our proofs of their optimality.
In Section~\ref{SO4-sec} we present the first such construction, which
proves that any two-qubit gate in $\mbox{\bf SO}(4)$ can be
implemented in 12 elementary (i.e., $R_{y}$, $R_{z}$) gates and 2
CNOTs. Theorem~\ref{O4-theorem} extends this results to any two-qubit gate in
$\mbox{\bf O}(4)$ with determinant equal to $-1$, and proves that
any such gate requires 12 elementary gates and 3 CNOTs. 
In Section~\ref{U4-sec} these results 
are generalized to the generic two-qubit gates in
$\mbox{\bf U}(4)$, and we provide an explicit construction that
requires 15 elementary gates and 3 CNOTs. Finally, 
in Section~\ref{cnot-sec} we prove that
our construction for generic two-qubit gates is optimal by showing
that there is at least one gate in $\mbox{\bf U}(4)$, namely the
two-qubit $\SWAP$ gate, which cannot be implemented in fewer than 3
CNOTs.

\section{Notation}
\label{notation-section}

Throughout this paper we identify a
quantum gate with the unitary matrix that defines its operation.
We take rotations about the $y$ and $z$-axes, respectively
$R_y(\theta)$ and $R_z(\alpha)$, as our elementary one-qubit
gates; i.e., 
\[ R_y(\theta) =\mbox{$\begin{pmatrix} 
     \cos\frac{\theta}{2} & \sin\frac{\theta}{2} \\
      & \\
     -\sin\frac{\theta}{2} & \cos\frac{\theta}{2}
     \end{pmatrix}$}, \quad
   R_z(\alpha) = \begin{pmatrix}
     e^{i\frac{\alpha}{2}} & 0 \\ 0 & e^{-i\frac{\alpha}{2}}
     \end{pmatrix}.\]
However, we also have three special one-qubit gates: 
the one-qubit identity matrix $\one_2$, and the Hadamard gate $H$ and
the phase gate $S$ defined as
\[ H = \mbox{$\frac{1}{\sqrt{2}}$} \begin{pmatrix} 1 & 1 \\ 1 & -1 \end{pmatrix}, \quad
   S = \begin{pmatrix} 1 & 0 \\ 0 & i \end{pmatrix}. \]
We define two CNOT gates, CNOT1 a standard CNOT gate
with the control on the top qubit and the target on the bottom
qubit, and CNOT2 with the control and target qubits flipped. Thus
\[   \mathrm{CNOT1}=\begin{pmatrix}
  1 & 0 & 0 & 0 \\ 0 & 1 & 0 & 0 \\ 0 & 0 & 0 & 1 \\ 0 & 0 & 1 & 0
 \end{pmatrix}, \ \mathrm{CNOT2}=\begin{pmatrix}
  1 & 0 & 0 & 0 \\ 0 & 0 & 0 & 1 \\ 0 & 0 & 1 & 0 \\ 0 & 1 & 0 & 0
 \end{pmatrix}. \]
We also use the two-qubit gate $\SWAP$ gate, which is defined as
\[ \SWAP=\mathrm{CNOT1}\cdot\mathrm{CNOT2}\cdot\mathrm{CNOT1} =
    \begin{pmatrix}
    1 & 0 & 0 & 0 \\ 0 & 0 & 1 & 0 \\
    0 & 1 & 0 & 0 \\ 0 & 0 & 0 & 1
    \end{pmatrix}. \]
We use the notation the $\wedge_1(V)$ for the controlled-$V$ gate,
where $V\in\mbox{\bf U}(2)$. Throughout this paper we assume that
for the $\wedge_1(V)$ gate the control qubit is the first (top)
qubit. Therefore,
\[  \wedge_1(V)=\begin{pmatrix} \one_2 & & \\ & & V \end{pmatrix}. \]
In the special case of the $\wedge_1(\sigma_z)$ gate, we use the
notation CZ. For any unitary matrix $U$, we denote its inverse, 
i.e., the conjugate-transpose of $U$, by $U^*$.

\section{Magic basis}
\label{magic-basis-sec}

\begin{figure}[!ht]
\thicklines
\begin{center}
\unitlength=.2mm
\begin{picture}(180,60)(0,10)
\drawline(0,0)(30,0)
\drawline(0,50)(30,50)
\drawline(30,35)(30,65)(60,65)(60,35)(30,35)
\drawline(30,-15)(30,15)(60,15)(60,-15)(30,-15)
\put(45,0){\makebox(0,0){$S$}}
\put(45,50){\makebox(0,0){$S$}}
\drawline(60,0)(90,0)
\drawline(90,-15)(90,15)(120,15)(120,-15)(90,-15)
\put(105,0){\makebox(0,0){$H$}}
\drawline(120,0)(180,0)
\drawline(60,50)(180,50)
\put(150,0){\circle*{5}}
\put(150,50){\circle{16}}
\drawline(150,0)(150,58)

\end{picture} \end{center}
\caption{A circuit for implementing the magic gate ${\cal M}$.}
\label{magic-fig}
\end{figure}
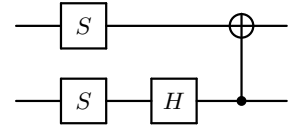

There are different ways to define the magic basis \cite{bennett,hill-wootters,kraus-cirac}.
Here we use the definition used in \cite{bennett,hill-wootters}:
\[ {\cal M}=\frac{1}{\sqrt{2}} \begin{pmatrix}
  1 & i & 0 & 0 \\ 0 & 0 & i & 1 \\ 0 & 0 & i & -1 \\ 1 & -i & 0 & 0
 \end{pmatrix} .  \]
The circuit of FIG.~\ref{magic-fig} implements this transformation.

The following theorem presents the basic property of the magic basis. 
This result is already known (see, e.g., {\cite{makhlin}), and we provide a 
proof for the sake of completeness.

\begin{theo}
For every real orthogonal matrix $U\in\mbox{\bf SO}(4)$, the matrix of $U$ in the
magic basis, i.e., ${\cal M}\cdot U\cdot {\cal M}^*$ is tensor product of two $2$-dimensional
special unitary matrices. In other words:
${\cal M}\cdot U\cdot {\cal M}^* \in \mbox{\bf SU}(2) \otimes \mbox{\bf SU}(2)$.
\label{magic-gate-theorem}
\end{theo}

\begin{proof}
{\em Proof.} We prove the theorem by showing that for every $A
\otimes B\in \mbox{\bf SU}(2) \otimes \mbox{\bf SU}(2)$, we have
${\cal M}^*\, \big(A \otimes B\big)\, {\cal M}\in \mbox{\bf
SO}(4)$. It is well-known that every matrix $A\in \mbox{\bf
SU}(2)$ can be written as the product $R_z(\alpha)\, R_y(\theta)\,
R_z(\beta)$, for some $\alpha, \beta$, and $\theta$. Therefore any
matrix $A \otimes B\in \mbox{\bf SU}(2) \otimes \mbox{\bf SU}(2)$
can be written as a product of the matrices of the form
$V\otimes\one_2$ and $\one_2\otimes V$, where $V$ is either
$R_y(\theta)$ or $R_z(\alpha)$. Thus the proof is complete if
${\cal M}^*\, \big(V\otimes\one_2\big)\, {\cal M}$ and ${\cal
M}^*\, \big(\one_2\otimes V\big)\, {\cal M}$, are in $\mbox{\bf
SO}(4)$. Elementary algebra shows that this the case.

Since the mapping $A\otimes B \mapsto {\cal M}^*\, \big( A\otimes B\big)\, {\cal M}$
is one-to-one and the spaces
$\mbox{\bf SU}(2) \otimes \mbox{\bf SU}(2)$ and  $\mbox{\bf SO}(4)$
have the same topological dimension, we conclude that this mapping
is an isomorphism between these two spaces. \qed
\end{proof}

\vspace{3mm}
Note that the above theorem is not true for all orthogonal matrices in
$\mbox{\bf O}(4)$. In fact, for every matrix $U \in \mbox{\bf O}(4)$, either
$\det (U)=1$ for which the above theorem holds, or $\det (U)=-1$ for
which we have the following theorem.

\begin{theo}
For every $U \in \mbox{\bf O}(4)$ with $\det (U)=-1$, the matrix
${\cal M}\, U\, {\cal M}^*$ is a tensor product of $2$-dimensional unitary matrices and
one $\SWAP$ gate in the form of the following decomposition:
$ {\cal M}\cdot U \cdot {\cal M}^* = \big( A\otimes B\big)\cdot \SWAP\cdot
   \big( \one_2\otimes \sigma_z\big), $
where $A,B \in \mbox{\bf U}(2)$.
\label{magic-gate-2-theorem}
\end{theo}

{\em Proof.}
First note that $\det(\mathrm{CNOT1})=-1$ and $\det(U\cdot\mathrm{CNOT1})=1$. Then
$ {\cal M}\, \big(\mathrm{CNOT1}\big)\, {\cal M}^*
   = \big({S}^*\otimes {S}^*\big)\, \SWAP\,
     \big(\one_2\otimes\sigma_z\big)$.
Since
${\cal M}\, U\, {\cal M}^*= \left({\cal M}\, \big(U \cdot \mathrm{CNOT1}\big) {\cal M}^*\right)
    \cdot \left( {\cal M}\, \big(\mathrm{CNOT1}\big)\, {\cal M}^* \right)$,
the theorem follows from Theorem~\ref{magic-gate-theorem}. \qed

\section{Realizing two-qubit gates from $\mbox{\bf O}(4)$}
\label{SO4-sec}
 
Let $U\in\mbox{\bf SO}(4)$. Then
Theorem~\ref{magic-gate-theorem} shows that ${\cal M}\, U\, {\cal
M}^*=A\otimes B$, where $A,B\in\mbox{\bf SU}(2)$. Therefore,
$U={\cal M}^*\, \big(A\otimes B)\, {\cal M}$. We use the circuit
of FIG.~\ref{magic-fig} for computing the magic basis transform
${\cal M}$ to obtain a circuit for computing the unitary operation
$U$. This circuit can be simplified by using the decompositions
$S=e^{i \pi/4} R_z(\pi/2)$ and $H=\sigma_z\,R_y(\pi/2)$. Note that
$\one_2\otimes\sigma_z$ and the CNOT2 gates commute, and the
overall phases $e^{i \pi/4}$ and  $e^{-i \pi/4}$ from $S$ and
$S^*$ cancel out. Hence we obtain the circuit of
FIG.~\ref{decomposition-1-fig} for computing a general two-qubit
gate from $\mbox{\bf SO}(4)$.
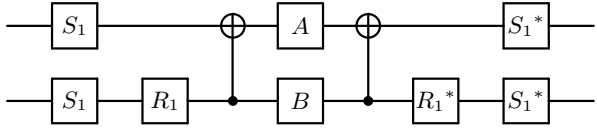
\begin{figure}[!ht]
\thicklines
\begin{center}
\unitlength=.2mm
\begin{picture}(390,60)(0,10)
\drawline(0,0)(30,0)
\drawline(0,50)(30,50)
\drawline(30,35)(30,65)(60,65)(60,35)(30,35)
\drawline(30,-15)(30,15)(60,15)(60,-15)(30,-15)
\put(45,0){\makebox(0,0){$S_1$}}
\put(45,50){\makebox(0,0){$S_1$}}
\drawline(60,0)(90,0)
\drawline(90,-15)(90,15)(120,15)(120,-15)(90,-15)
\put(105,0){\makebox(0,0){$R_1$}}
\drawline(120,0)(180,0)
\drawline(60,50)(180,50)
\put(150,0){\circle*{5}}
\put(150,50){\circle{16}}
\drawline(150,0)(150,58)
\drawline(180,-15)(180,15)(210,15)(210,-15)(180,-15)
\drawline(180,35)(180,65)(210,65)(210,35)(180,35)
\put(195,0){\makebox(0,0){$B$}}
\put(195,50){\makebox(0,0){$A$}}
\drawline(210,0)(270,0)
\drawline(210,50)(330,50)
\put(240,0){\circle*{5}}
\put(240,50){\circle{16}}
\drawline(240,0)(240,58)
\drawline(270,-15)(270,15)(300,15)(300,-15)(270,-15)
\put(285,0){\makebox(0,0){${R_1}^*$}}
\drawline(300,0)(330,0)
\drawline(330,35)(330,65)(360,65)(360,35)(330,35)
\drawline(330,-15)(330,15)(360,15)(360,-15)(330,-15)
\put(345,0){\makebox(0,0){${S_1}^*$}}
\put(345,50){\makebox(0,0){${S_1}^*$}}
\drawline(360,0)(390,0)
\drawline(360,50)(390,50)

\end{picture} \end{center}
\caption{A circuit for implementing a general transform in $\mbox{\bf SO}(4)$, where
       $A,B\in\mbox{\bf SU}(2)$, $S_1=R_z(\pi/2)$ and $R_1=R_y(\pi/2)$.}
\label{decomposition-1-fig}
\end{figure}
\noindent
Thus we have proved the following theorem.

\begin{theo}
Every two-qubit quantum gate in $\mbox{\bf SO}(4)$ can be realized by a circuit
consisting of $12$ elementary one-qubit gates and $2$ {\em CNOT} gates.
\label{SO4-theorem}
\end{theo}

A similar argument and Theorem~\ref{magic-gate-2-theorem} imply
the following construction for gates from $\mbox{\bf O}(4)$ with
determinant equal to $-1$.

\begin{theo}
Every two-qubit quantum gate in $\mbox{\bf O}(4)$ determinant equal to $-1$
can be realized by a circuit
consisting of\/ $12$ elementary gates and\/ $2$ {\em CNOT} gates and one {\em $\SWAP$} gate
(see {\em FIG.~\ref{decomposition-2-fig}}).
\label{O4-theorem}
\end{theo}

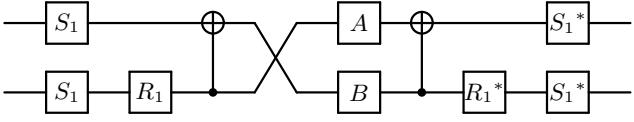
\begin{figure}[!ht]
\thicklines
\begin{center}
\unitlength=.185mm
\begin{picture}(450,60)(0,10)
\drawline(0,0)(30,0)
\drawline(0,50)(30,50)
\drawline(30,35)(30,65)(60,65)(60,35)(30,35)
\drawline(30,-15)(30,15)(60,15)(60,-15)(30,-15)
\put(45,0){\makebox(0,0){$S_1$}}
\put(45,50){\makebox(0,0){$S_1$}}
\drawline(60,0)(90,0)
\drawline(90,-15)(90,15)(120,15)(120,-15)(90,-15)
\put(105,0){\makebox(0,0){$R_1$}}
\drawline(120,0)(180,0)
\drawline(60,50)(180,50)
\put(150,0){\circle*{5}}
\put(150,50){\circle{16}}
\drawline(150,0)(150,58)
\drawline(180,0)(210,50)
\drawline(180,50)(210,0)
\drawline(210,0)(240,0)
\drawline(210,50)(240,50)

\drawline(240,-15)(240,15)(270,15)(270,-15)(240,-15)
\drawline(240,35)(240,65)(270,65)(270,35)(240,35)
\put(255,0){\makebox(0,0){$B$}}
\put(255,50){\makebox(0,0){$A$}}
\drawline(270,0)(330,0)
\drawline(270,50)(390,50)
\put(300,0){\circle*{5}}
\put(300,50){\circle{16}}
\drawline(300,0)(300,58)
\drawline(330,-15)(330,15)(360,15)(360,-15)(330,-15)
\put(345,0){\makebox(0,0){${R_1}^*$}}
\drawline(360,0)(390,0)
\drawline(390,35)(390,65)(420,65)(420,35)(390,35)
\drawline(390,-15)(390,15)(420,15)(420,-15)(390,-15)
\put(405,0){\makebox(0,0){${S_1}^*$}}
\put(405,50){\makebox(0,0){${S_1}^*$}}
\drawline(420,0)(450,0)
\drawline(420,50)(450,50)

\end{picture} \end{center}
\caption{A circuit for implementing a transform in $\mbox{\bf O}(4)$ determinant equal to $-1$, 
        where $A,B\in\mbox{\bf SU}(2)$, $S_1=R_z(\frac{\pi}{2})$ and $R_1=R_y(\frac{\pi}{2})$.}
\label{decomposition-2-fig}
\end{figure}

\noindent Next, we generalize these results to construct circuits
for gates in $\mbox{\bf U}(4)$.

\section{Realizing two-qubit gates from $\mbox{\bf U}(4)$}
\label{U4-sec}

In is known that every $U\in\mbox{\bf U}(4)$ can be written as
\begin{equation}
 U = \big( A_1 \otimes A_2\big)\cdot N(\alpha,\beta,\gamma) \cdot
       \big( A_3 \otimes A_4\big), 
\label{cartan-equ}
\end{equation}
where $A_j\in\mbox{\bf U}(2)$ and
\[ N(\alpha,\beta,\gamma) = \big[ \exp\big( i(\alpha\, \sigma_x\otimes\sigma_x +\beta\, 
    \sigma_y\otimes\sigma_y
           +\gamma\, \sigma_z\otimes\sigma_z) \big) \big], \]          
for $\alpha,\beta,\gamma\in\rr$ (see, e.g., \cite{khaneja,kraus-cirac,zhang}).
Note that if $U\in\mbox{\bf SU}(4)$, then we can choose all operations $A_j$ 
in (\ref{cartan-equ}) from $\mbox{\bf SU}(2)$.
Our construction is based on constructing an optimal circuit
for computing $N(\alpha,\beta,\gamma)$.
To this end, we first note that
$ D = {\cal M}^*\cdot N\cdot {\cal M}$ is a diagonal matrix of the form
\[ \mathrm{diag}\left(
      e^{i(\alpha-\beta+\gamma)}, e^{-i(\alpha-\beta-\gamma)},
      e^{i(\alpha+\beta-\gamma)}, e^{-i(\alpha+\beta+\gamma)}\right). \]
Therefore,
$ N(\alpha,\beta,\gamma)= {\cal M} \cdot D \cdot {\cal M}^*$.
Utilizing the circuit of FIG.~\ref{magic-fig} for ${\cal M}$, we get the
circuit of FIG.~\ref{decomposition-3-fig} for computing $N(\alpha,\beta,\gamma)$.
Note that
$\big( S\otimes S)\cdot D \cdot \big(S^*\otimes S^*\big)=D$.
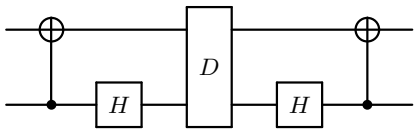
\begin{figure}[!ht]
\thicklines
\begin{center}
\unitlength=.2mm
\begin{picture}(270,65)(0,10)

\drawline(0,0)(60,0)
\drawline(0,50)(120,50)
\put(30,0){\circle*{5}}
\put(30,50){\circle{16}}
\drawline(30,0)(30,58)
\drawline(60,-15)(60,15)(90,15)(90,-15)(60,-15)
\put(75,0){\makebox(0,0){$H$}}
\drawline(90,0)(120,0)
\drawline(120,-15)(120,65)(150,65)(150,-15)(120,-15)
\put(135,25){\makebox(0,0){$D$}}
\drawline(150,0)(180,0)
\drawline(150,50)(270,50)
\drawline(180,-15)(180,15)(210,15)(210,-15)(180,-15)
\put(195,0){\makebox(0,0){$H$}}
\drawline(210,0)(270,0)
\put(240,0){\circle*{5}}
\put(240,50){\circle{16}}
\drawline(240,0)(240,58)

\end{picture} \end{center}
\caption{A circuit for implementing $N(\alpha,\beta,\gamma)$; first version.}
\label{decomposition-3-fig}
\end{figure}
Then we substitute the right-hand side Hadamard gate of FIG.~\ref{decomposition-3-fig}
by $3$ gates, using the following identity:
$\one_2\otimes H = \mathrm{CNOT1}\cdot \big( \one_2\otimes H \big)\cdot \mathrm{CZ}$.
Now, the matrix $D_1=\mathrm{CZ}\cdot D$ is a diagonal matrix, and 
\begin{equation}
  (\one_2\otimes H)\cdot D_1\cdot (\one_2\otimes H) =
   \Lambda_1(V_2)\cdot (\one_2\otimes V_1),
\label{D1-equ}
\end{equation}
where 
\begin{align*}
 V_1 &= \begin{pmatrix} 
     e^{i\, \gamma}\cos(\alpha-\beta) & i\,e^{i\, \gamma}\sin(\alpha-\beta) \\
     i\,e^{i\, \gamma}\sin(\alpha-\beta) & e^{i\, \gamma}\cos(\alpha-\beta)
     \end{pmatrix}, \\
 V_2 &= \begin{pmatrix} 
     i\, e^{-2i\, \gamma}\sin 2\beta & e^{-2i\, \gamma}\cos 2\beta \\
     e^{-2i\, \gamma}\cos 2\beta & i\, e^{-2i\, \gamma}\sin 2\beta
     \end{pmatrix}.
\end{align*}
We have the following decompositions for $V_1$ and $\Lambda_1(V_2)$
(see also \cite{cybenko}):
\begin{equation}
 V_1 = e^{i\, \gamma} R_z(-{\textstyle \frac{\pi}{2}})\cdot R_y(2(\beta-\alpha))
     \cdot R_z({\textstyle \frac{\pi}{2}}), 
\label{V1-equ}
\end{equation}
and
\begin{equation}
\begin{split}
\Lambda_1(V_2) &= e^{i\, ({\textstyle \frac{\pi}{4}}-\gamma)} 
   \big(\one_2\otimes R_z(-{\textstyle \frac{\pi}{2}})\big)\cdot \mathrm{CNOT1}  \\
   &\phantom{=} \cdot \big(\one_2\otimes R_y(2\beta-{\textstyle \frac{\pi}{2}})\big) 
                \cdot \mathrm{CNOT1}\\
   &\phantom{=} \cdot 
  \big(R_z(2\gamma-{\textstyle \frac{\pi}{2}})\otimes (R_y({\textstyle \frac{\pi}{2}}-2\beta)\cdot
  R_z({\textstyle \frac{\pi}{2}}))\big).
\end{split}
\label{V2-equ}
\end{equation}
By utilizing the equations (\ref{D1-equ})--(\ref{V2-equ}), we can convert the circuit of
FIG.~\ref{decomposition-3-fig} to the circuit of FIG.~\ref{decomposition-4-fig}
\begin{figure}[!ht]
\thicklines
\begin{center}
\unitlength=.19mm
\begin{picture}(420,65)(0,10)

\drawline(0,0)(60,0)
\drawline(0,50)(120,50)
\drawline(120,35)(120,65)(150,65)(150,35)(120,35)
\put(135,50){\makebox(0,0){$S_2$}}
\drawline(150,50)(420,50)
\put(30,0){\circle*{5}}
\put(30,50){\circle{16}}
\drawline(30,0)(30,58)
\drawline(60,-15)(60,15)(90,15)(90,-15)(60,-15)
\put(75,0){\makebox(0,0){$S_1$}}
\drawline(90,0)(120,0)
\drawline(120,-15)(120,15)(150,15)(150,-15)(120,-15)
\put(135,0){\makebox(0,0){$T_1$}}
\drawline(150,0)(210,0)
\put(180,50){\circle*{5}}
\put(180,0){\circle{16}}
\drawline(180,50)(180,-8)
\drawline(210,-15)(210,15)(240,15)(240,-15)(210,-15)
\put(225,0){\makebox(0,0){$T_2$}}
\drawline(240,0)(300,0)
\put(270,50){\circle*{5}}
\put(270,0){\circle{16}}
\drawline(270,50)(270,-8)
\drawline(300,-15)(300,15)(330,15)(330,-15)(300,-15)
\put(315,0){\makebox(0,0){${S_1}^*$}}
\drawline(330,0)(420,0)
\put(360,50){\circle*{5}}
\put(360,0){\circle{16}}
\drawline(360,50)(360,-8)
\put(390,0){\circle*{5}}
\put(390,50){\circle{16}}
\drawline(390,0)(390,58)

\end{picture} \end{center}
\caption{A circuit for implementing $N(\alpha,\beta,\gamma)$; second version.
Here $S_1=R_z({\textstyle \frac{\pi}{2}})$, $S_2=R_z(2\gamma-{\textstyle \frac{\pi}{2}})$,
$T_1=R_y({\textstyle \frac{\pi}{2}}-2\alpha)$, and $T_2=R_y({\textstyle 2\beta-\frac{\pi}{2}})$. }
\label{decomposition-4-fig}
\end{figure}
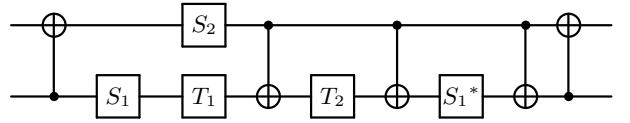

Now we focus on the sequence
$\mathrm{CNOT1} \cdot  \big( \one_2\otimes R_z(-\frac{\pi}{2})\big) \cdot\mathrm{CNOT1}$
of operations. We have the following identity
\begin{multline*}
 \mathrm{CNOT1} \cdot  \big( \one_2\otimes R_z(\theta)\big) \cdot\mathrm{CNOT1} = \\
   \mathrm{CNOT2} \cdot  \big(R_z(\theta) \otimes \one_2\big) \cdot\mathrm{CNOT2} .
\end{multline*}
After applying this rule, the two consecutive CNOT2 gates on the right-hand side of the
circuit reduce to the identity. 
Also note that, on the left-hand side of the circuit, we can apply the rule
\[ \big( \one_2\otimes R_z(\theta)\big) \cdot \mathrm{CNOT1} = 
   \mathrm{CNOT1} \cdot \big( \one_2\otimes R_z(\theta)\big) . \]
Thus the circuit of FIG.~\ref{decomposition-4-fig} can be converted to the circuit of
FIG.~\ref{decomposition-5-fig}. Note that the operation defined by this circuit has 
determinant equal to $-1$, thus we need to add a global $e^{i\, \frac{\pi}{4}}$ phase 
to get the special unitary operation $N(\alpha,\beta,\gamma)$ exactly.
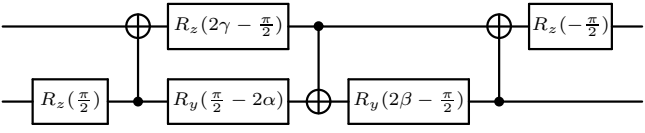
\begin{figure}[!ht]
\thicklines
\begin{center}
\unitlength=.2mm
\begin{picture}(425,65)(0,10)

\drawline(0,0)(20,0)
\drawline(0,50)(110,50)
\drawline(20,-15)(20,15)(70,15)(70,-15)(20,-15)
\put(45,0){\makebox(0,0){\scriptsize $R_z({\textstyle \frac{\pi}{2}})$}}
\drawline(70,0)(110,0)
\put(90,0){\circle*{5}}
\put(90,50){\circle{16}}
\drawline(90,0)(90,58)
\drawline(110,-15)(110,15)(190,15)(190,-15)(110,-15)
\put(150,0){\makebox(0,0){\scriptsize $R_y({\textstyle \frac{\pi}{2}}-2\alpha)$}}
\drawline(110,35)(110,65)(190,65)(190,35)(110,35)
\put(150,50){\makebox(0,0){\scriptsize $R_z(2\gamma-{\textstyle \frac{\pi}{2}})$}}
\drawline(190,0)(230,0)\drawline(310,0)(425,0)
\drawline(190,50)(350,50)\drawline(405,50)(425,50)
\put(210,50){\circle*{5}}
\put(210,0){\circle{16}}
\drawline(210,50)(210,-8)
\drawline(230,-15)(230,15)(310,15)(310,-15)(230,-15)
\put(270,0){\makebox(0,0){\scriptsize $R_y({\textstyle 2\beta-\frac{\pi}{2}})$}}
\put(330,0){\circle*{5}}
\put(330,50){\circle{16}}
\drawline(330,0)(330,58)

\drawline(350,35)(350,65)(405,65)(405,35)(350,35)
\put(377.5,50){\makebox(0,0){\scriptsize $R_z({\textstyle -\frac{\pi}{2}})$}}

\end{picture} \end{center}
\caption{A circuit for implementing $N(\alpha,\beta,\gamma)$; third version.
A global $e^{i\, \frac{\pi}{4}}$ phase is  missing here.}
\label{decomposition-5-fig}
\end{figure}
Now utilizing the circuit of FIG.~\ref{decomposition-5-fig} and the canonical 
decomposition (\ref{cartan-equ}), we could get a circuit to realize the 
operation $U\in\mbox{\bf U}(4)$. Note that in this process, the left and right-hand
side operations $R_z(\frac{\pi}{2})$ and $R_z(-\frac{\pi}{2})$ of 
FIG.~\ref{decomposition-5-fig} will be ``absorbed'' by adjacent $A_j$.
The final result is the circuit of FIG.~\ref{decomposition-6-fig},
and we have proved the following theorem.

\begin{theo}
Every two-qubit quantum gate in $\mbox{\bf U}(4)$ can be realized, up to a global phase, 
by a circuit
consisting of\/ $15$\/ elementary one-qubit gates and $3$ {\em CNOT} gates.
\label{U4-theorem}
\end{theo}

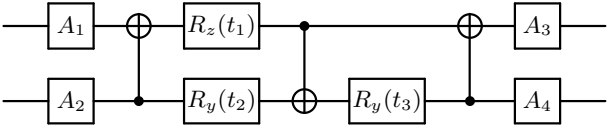
\begin{figure}[!ht]
\thicklines
\begin{center}
\unitlength=.2mm
\begin{picture}(400,65)(0,10)
\drawline(0,0)(30,0)
\drawline(0,50)(30,50)
\drawline(30,35)(30,65)(60,65)(60,35)(30,35)
\drawline(30,-15)(30,15)(60,15)(60,-15)(30,-15)
\put(45,0){\makebox(0,0){$A_2$}}
\put(45,50){\makebox(0,0){$A_1$}}
\drawline(60,0)(120,0)
\drawline(60,50)(120,50)
\put(90,0){\circle*{5}}
\put(90,50){\circle{16}}
\drawline(90,0)(90,58)
\drawline(120,-15)(120,15)(170,15)(170,-15)(120,-15)
\put(145,0){\makebox(0,0){\small $R_y(t_2)$}}
\drawline(120,35)(120,65)(170,65)(170,35)(120,35)
\put(145,50){\makebox(0,0){\small $R_z(t_1)$}}
\drawline(170,0)(230,0)
\drawline(170,50)(340,50)
\put(200,50){\circle*{5}}
\put(200,0){\circle{16}}
\drawline(200,50)(200,-8)
\drawline(230,-15)(230,15)(280,15)(280,-15)(230,-15)
\put(255,0){\makebox(0,0){\small $R_y(t_3)$}}
\drawline(280,0)(340,0)
\put(310,0){\circle*{5}}
\put(310,50){\circle{16}}
\drawline(310,0)(310,58)
\drawline(340,35)(340,65)(370,65)(370,35)(340,35)
\drawline(340,-15)(340,15)(370,15)(370,-15)(340,-15)
\put(355,0){\makebox(0,0){$A_4$}}
\put(355,50){\makebox(0,0){$A_3$}}
\drawline(370,0)(400,0)
\drawline(370,50)(400,50)

\end{picture} \end{center}
\caption{A circuit for implementing a transform in $\mbox{\bf U}(4)$.}
\label{decomposition-6-fig}
\end{figure}
The construction given in Theorem~\ref{U4-theorem} is {\em optimal}. To prove
this it is sufficient to place a lower bound on the number of CNOT
gates needed to implement a generic two-qubit gate. This is
because \cite{bullock-markov} already shows that we need at least
15 elementary one-qubit gates, to implement a generic two-qubit
gate. So we need only concern ourselves with the minimum required
number of CNOT gates. We prove in the next section that three CNOT gates are needed 
in the general case. 

We wish to emphasize that our decomposition is {\em
constructive}. To see this, note that we can use Kraus and
Cirac's methods \cite{kraus-cirac} to decompose any desired two-qubit gate
into the form given by equation (\ref{cartan-equ}). All parameters in this
decomposition may be determined constructively. Thereafter,
it only remains to reduce the $N(\alpha,\beta,\gamma)$ matrix
to an explicit quantum circuit. This we can do immediately
using the circuit template in FIG.~\ref{decomposition-5-fig}. 
By concatenating these
two processes we can find the optimal circuit for any
generic two-qubit operation constructively

\section{Three CNOT gates are needed}
\label{cnot-sec}

To show that the
construction of Theorem~\ref{U4-theorem} is optimal, we prove that
there is at least one gate in $\mbox{\bf U}(4)$, namely the
two-qubit $\SWAP$ gate, a real unitary matrix having a determinant of
$-1$, which requires no less than $3$ CNOT gates. 

In the proof of the following theorem we utilize the notion of 
{\em entangling power} introduced in \cite{zanardi}.
For a unitary operation $U\in\mbox{\bf U}(4)$, the entangling power of
$U$ is defined as
\[ \EP(U) = \mathop{\mathrm {average}}_{\ket{\psi_1}\otimes \ket{\psi_2}} 
    \big[ E \big( U\ket{\psi_1}\otimes \ket{\psi_2}\big) \big], \]
where average is over all product states 
$\ket{\psi_1}\otimes \ket{\psi_2}\in \cc^2\otimes \cc^2$ distributed
according to the uniform distribution (in general, we can define $\EP$ with 
regards to any distribution, but here we only consider the uniform distribution).
In the above formula
$E$ is the linear entropy {\em entanglement measure} defined for $\ket{\psi}\in\cc^4$ 
as follows:
\[  E\big(\ket{\psi}\big) = 1-\mathrm{tr}_1\, \rho^2, \]   
where $\rho=\mathrm{tr}_2\,\project{\psi}{\psi}$ and $\mathrm{tr}_j$ denotes the
result of tracing out the $j^{\mathrm{th}}$ qubit.
Note that $0\leq E\big(\ket{\psi}\big) \leq \frac{3}{4}$, and the lower or upper 
bound is obtained if $\ket{\psi}$ is a product state or a maximally entangled
state, respectively.
In \cite{zanardi} the following simple formula for calculating $\EP$ is presented:
\begin{multline*}
  \EP(U) = \textstyle{\frac{5}{9}} - \textstyle{\frac{1}{36}}\left[
     \mean{U^{\otimes 2}, T_{1,3}\,U^{\otimes 2}\,T_{1,3}} + \right. \\
    \left.  \mean{(\SWAP\cdot U)^{\otimes 2}, 
    T_{1,3}\,(\SWAP\cdot U)^{\otimes 2}\,T_{1,3}}
     \right],
\end{multline*}
where the Hilbert-Schmidt scalar product $\mean{A,B}$ is defined as 
$\mean{A,B} = \mathrm{tr} (A^\dagger B)$ and 
the permutation $T_{1,3}$ on $\cc^2\otimes\cc^2\otimes\cc^2\otimes\cc^2$
is the transposition $T_{1,3}\ket{a,b,c,d}=\ket{c,b,a,d}$ on the system
of 4 qubits.  

We will utilize the following basic properties of the function $\EP$.
\begin{itemize}

\item 
For every $U\in\mbox{\bf U}(4)$ we have
$0\leq \EP \big( U \big) \leq \frac{2}{9}$. 

\item 
For every $A,B\in\mbox{\bf U}(2)$ we have $\EP(A\otimes B)=0$.

\item
For every $U\in\mbox{\bf U}(4)$ and $A,B\in\mbox{\bf U}(2)$ we have
$\EP\big((A\otimes B)\cdot U\big)=\EP\big(U\cdot(A\otimes B)\big)=\EP(U)$.

\item
$\EP(U)=\EP(U^*)$.

\item
$\EP(\mathrm{CNOT}) = \frac{2}{9}$ and $\EP(\SWAP) = 0$.

\end{itemize}
We will also use the simple fact that $\SWAP$ cannot be written as
$\SWAP=A\otimes B$, where $A,B\in\mbox{\bf U}(2)$.

\begin{theo}
To compute the $\SWAP$ at least $3$ {\em CNOT} gates are needed.
\label{3cnot-theorem}
\end{theo}

{\em Proof.} We construct a proof by contradiction. Suppose that
there is a circuit computing $\SWAP$ and consists of less than three
CNOT gates. We consider two possible cases.

\vspace{4mm}
\noindent
{\em Case 1.}
Suppose that $\SWAP$ is computed by a circuit consisting of two CNOT gates.
We substitute each CNOT gate by a small subcircuit in terms of CZ 
(controlled-$\sigma_z$) gate; i.e., 
\[ \mathrm{CNOT}=\big(\one_2\otimes H\big)\cdot \mathrm{CZ} \cdot \big(\one_2\otimes H\big).\]
Then by utilizing the following commutation rules
\begin{align*}
 \mathrm{CZ} \cdot \big(\one_2\otimes R_z(t)\big) 
    &= \big(\one_2\otimes R_z(t)\big) \cdot \mathrm{CZ}, \\
 \mathrm{CZ} \cdot \big(R_z(t)\otimes \one_2\big) 
    &= \big(R_z(t)\otimes \one_2\big) \cdot \mathrm{CZ},   
\end{align*}
we obtain the simplified circuit of FIG.~\ref{CZ-fig} for computing the $\SWAP$ gate.
\begin{figure}[!ht]
\thicklines
\begin{center}
\unitlength=.2mm
\begin{picture}(350,65)(0,10)
\drawline(0,0)(30,0)
\drawline(0,50)(30,50)
\drawline(30,35)(30,65)(60,65)(60,35)(30,35)
\drawline(30,-15)(30,15)(60,15)(60,-15)(30,-15)
\put(45,0){\makebox(0,0){$A_2$}}
\put(45,50){\makebox(0,0){$A_1$}}
\drawline(60,0)(90,0)
\drawline(60,50)(150,50)
\drawline(90,-15)(90,15)(120,15)(120,-15)(90,-15)
\put(105,0){\makebox(0,0){$\sigma_z$}}
\put(105,50){\circle*{5}}
\drawline(105,50)(105,15)
\drawline(120,0)(150,0)
\drawline(150,35)(150,65)(200,65)(200,35)(150,35)
\drawline(150,-15)(150,15)(200,15)(200,-15)(150,-15)
\put(175,0){\makebox(0,0){$R_y(b)$}}
\put(175,50){\makebox(0,0){$R_y(a)$}}
\drawline(200,0)(230,0)
\drawline(200,50)(290,50)
\drawline(230,-15)(230,15)(260,15)(260,-15)(230,-15)
\put(245,0){\makebox(0,0){$\sigma_z$}}
\put(245,50){\circle*{5}}
\drawline(245,50)(245,15)
\drawline(260,0)(290,0)
\drawline(290,35)(290,65)(320,65)(320,35)(290,35)
\drawline(290,-15)(290,15)(320,15)(320,-15)(290,-15)
\put(305,0){\makebox(0,0){$A_4$}}
\put(305,50){\makebox(0,0){$A_3$}}
\drawline(320,0)(350,0)
\drawline(320,50)(350,50)

\end{picture} \end{center}
\caption{A circuit consisting of two CNOT gates in terms of CZ gates.}
\label{CZ-fig}
\end{figure}
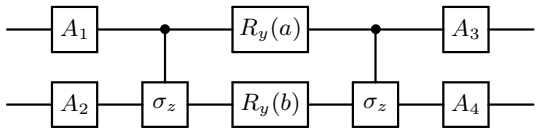
Note that in this figure we choose the top (first) qubit as the control qubit for
the CZ gates, but we could choose the other qubit as the control qubit as well, since the
action of the CZ gate is not change by switching the control and target qubits. Now, let
\[ U = \mathrm{CZ} \cdot \big( R_y(a)\otimes R_y(b) \big) \cdot \mathrm{CZ}. \]
Then 
\begin{multline*}
 \EP(U)=\EP(\SWAP)  \\
   ={\textstyle \frac{1}{18}} 
     \big( 3-\cos(2a)-\cos(2b)-\cos(2a)\cos(2b) \big)=0.
\end{multline*}
Therefore, $a,b\in\{0,\pi\}$. Thus we have the following four possible cases for the
unitary operation $U$:
\begin{itemize}

\item if $a=b=0$, then $U=\one_2$;

\item if $a=0$, $b=\pi$, then $U=\sigma_z\otimes R_y(\pi)$;

\item if $a=\pi$, $b=0$, then $U=R_y(\pi)\otimes \sigma_z$;

\item if $a=b=\pi$, then $U=-\sigma_x\otimes \sigma_x$.

\end{itemize}
In each case, we conclude that $\SWAP=V_1\otimes V_2$, for some
$V_1,V_2\in\mbox{\bf U}(2)$, which is a contradiction.

\vspace{4mm}
\noindent
{\em Case 2.}
Suppose that $\SWAP$ is computed by a circuit consisting of only one CNOT gat; 
for example
\[   \SWAP =  \big(A_1 \otimes A_2\big)\cdot \mathrm{CNOT1} \cdot
                    \big(A_3 \otimes A_4\big), \]
where $A_j\in\mbox{\bf U}(2)$.
Then $\EP(\SWAP)=\EP(\mathrm{CNOT})$, which again is a contradiction. \qed

\section{Conclusion}

In this paper we prove tight bounds on the numbers of
one-qubit gates and CNOT gates needed to implement generic
two-qubit quantum computations. In addition, we give a
constructive procedure for finding such decompositions,
which uses the Kraus-Cirac decomposition to find the core
entangling operation underlying the two-qubit gate, i.e.,
$N(\alpha, \beta, \gamma)$, and then substitutes the discovered
parameter values into an equivalent circuit template for
$N(\alpha, \beta, \gamma)$ as shown in FIG.~\ref{decomposition-5-fig}. 
The net result is
an explicit circuit for any desired two-qubit unitary
operation that uses at most three CNOTs and 15 elementary $y$-
or $z$- single qubit rotations.

We point out that it is possible to decompose a desired
unitary operation into many different families of quantum
gates. For example, the basis of all one-qubit gates augmented
with CNOT was first studied in \cite{barenco}, and was shown
to be capable of implementing any $n$-qubit unitary
operation {\em exactly}. This scheme has the advantage that
only a single, fixed, type of two-qubit gate need be built.
Similar schemes are known that use different fixed
entangling operations such as $i$$\SWAP$ gates (in
superconducting quantum computing) and $\sqrt{\SWAP}$ gates
(in spintronic quantum computing). In addition, other
decompositions are possible that use parameterized two-qubit
gates. These may lead to more efficient factorizations in
special cases, but also make for a more complicated quantum
computer architecture.

The motivation for our work comes from the fact that it is
still very difficult, experimentally, to implement multiple
quantum gates. Thus, in order to attain near term
experimental milestones, it will be important to minimize
the number of gates
they require. Although our scheme yields minimal circuits
for generic two qubit operations, further reductions are
still possible in certain special cases. We therefore
augment our procedure with rewrite rules, to find
even simpler circuits if they exist. Hence, our new
construction brings certain state synthesis tasks within the
grasp of experimentalists.

In addition, as quantum circuits for (arbitrary) $n$-qubit
operations are always expressed in terms of a sequence of
one-qubit and two-qubit gates, by designing component two-qubit
operations minimally, we can sometimes improve the
efficiency of implementing $n$-qubit computations.

\begin{acknowledgments}
The research described in this
paper was performed at the Jet Propulsion Laboratory (JPL),
California Institute of Technology, under contract with National
Aeronautics and Space Administration (NASA). We would like to
thank the sponsors, the National Security Agency (NSA), and the
Advanced Research and Development Activity (ARDA), for their
support.
\end{acknowledgments}

\end{document}